\documentclass[aps,prb,multicol,epsf]{revtex4}
\usepackage{graphicx}
\usepackage{epsfig}

\newcommand{\ba}{\begin{eqnarray}} 
\newcommand{\ea}{\end{eqnarray}} 
\newcommand{\be}{\begin{equation}} 
\newcommand{\ee}{\end{equation}} 
\newcommand{\bea}{\begin{eqnarray}} 
\newcommand{\eea}{\end{eqnarray}} 

\def\etal{{\it et al}.}

\def\eqnref#1{Eq.(\ref{#1})}
\def\sectref#1{Section \ref{#1}}
\def\tblref#1{Table \ref{#1}}
\def\figref#1{Figure \ref{#1}}

\renewcommand{\baselinestretch}{2}
\begin{document}

\def\CC{{\rm\kern.24em \vrule width.04em height1.46ex depth-.07ex \kern-.30em C}
}

\title{A Simple Model for Magnetization Ratios in Doped Nanocrystals}

\author{Joshua Schrier and K. Birgitta Whaley}

\affiliation{Department of Chemistry and Pitzer Center for Theoretical
Chemistry, University of California, Berkeley 94720}

\begin{abstract}

Recent experiments on Mn-doped ZnS nanocrystals have shown unusual
magnetization properties.  We describe a nearest-neighbor Heisenberg
exchange model for calculating the magnetization ratios of these
antiferromagnetically doped crystals, in which the dopant atoms are
distributed inhomogeneously within the nanocrystal.  This simple
inhomogeneous doping model is capable of reproducing the experimental
results, and suggests that interior dopant atoms are localized within
the crystal.

\end{abstract}

\maketitle

\section{Introduction} \label{intro}

Transition metal doping is now possible for a variety of
nanocrystalline semiconductor materials, e.g., CdS,\cite{CGB98,FLIP99}
CdSe,\cite{MKB+00,HMS02}, ZnSe,\cite{NYCK01} and
ZnS.\cite{YHR03,TKK03} In bulk dilute magnetic semiconductors (DMS),
Heisenberg coupling of magnetic dopants, such as Mn, occurs via
anion-mediated superexchange,\cite{SLT+86, Furd88} but it is not
completely understood how this is modulated by finite nanocrystal (NC) size
effects.\cite{FLIP99}

In a recent experiment, Tsuji \etal\, synthesized Mn-doped ZnS
NCs and determined the ratio between the magnetization at a finite
field and the expected magnetization for the given number of dopants
per NC, defined as $M_{5T}/M_{sat}$.\cite{TKK03}  The initially
prepared NCs (designated ``as prepared'') had a 1.2\% Mn dopant
concentration, an average of 4.3 Mn dopants per NC (mostly on
the surface), and $M_{5T}/M_{sat}=91\%$.  Washing the sample
with acid removed the surface atoms, yielding a sample (designated
``HCl-washed'') with a 0.10\% Mn concentration, an average of 0.3-0.5
Mn dopants per NC (presumably on the interior), and
$M_{5T}/M_{sat}=77\%$.  The effect of antiferromagnetic coupling of the Mn ions
is to reduce the magnetization ratio. However,
the ``as prepared'' NCs, with their higher
dopant concentration (and hence increased chance of exchange
interaction between dopants), show a higher $M_{5T}/M_{sat}$ than 
the ``HCl-washed'' samples, with their order of magnitude lower dopant concentration.
Tsuji \etal\, posit that Mn ions on the surface do not interact
magnetically.  However, in this paper we show that even with uniform
Heisenberg exchange constants for surface and interior atoms, 
taking into account the nonhomogeneous distribution of dopant atoms within the
NC alone is sufficient to explain the magnetization results.

\section{Model}
\label{model}
\begin{figure}
\includegraphics[width=2.75in]{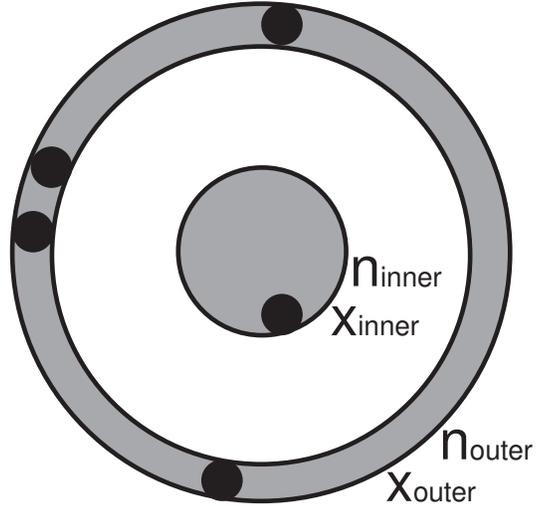}
\caption{Schematics of the inhomogeneous dopant distribution model
proposed in this paper.  $x_{inner}$ and $x_{outer}$ are the dopant
concentrations for the inner and outer shells respectively.
$n_{inner}$ and $n_{outer}$ are the number of
cation sites enclosed within the inner and outer shells, respectively.
Black circles show a random configuration of dopant atoms.  For the
low dopant concentration (``HCl washed'') samples, the restriction of
doping to a small inner shell increases the probability of pairs as
compared to the case where the dopant atoms are uniformly distributed
over the entire NC, hence reducing $M_{5T}/M_{sat}$.  For high
dopant concentration (``as prepared'') samples, $x_{outer} >
x_{inner}$ reduces the probability of pairing between surface and
interior dopant atoms 
as compared with
the uniformly distributed case,
hence increasing $M_{5T}/M_{sat}$ .}
\label{inhomogeneousFig}
\end{figure}

The experiments of Tsuji \etal\, are conducted with 3 nm diameter ZnS
NCs.\cite{TKK03} Using the zinc-blende model of Lippens and Lanoo, this
corresponds to a 729 atom NC, with alternating ``shells'' of
anion and cation atoms, summarized in \tblref{shells}.\cite{LL89}
Assuming random incorporation of dopant atoms during the growth
process, the distribution of dopant atoms within the crystals obeys a
binomial law as noted by Counio \etal,\cite{CGB98} where the
probability, $p(N_{d})$, that a NC contains $N_{d}$
dopant atoms, is given by
\be\label{prob_law}
p(N_{d}) = \left ( \matrix{ 
n_{sites} \cr
N_{d} \cr}
\right ) x^{N_{d}} {\left ( 1- x \right )}^{n_{sites}-N_{d}}
\ee
where
\be
\left ( \matrix{
n \cr
m \cr} \right )
\ee
is a binomial coefficient, $x$ is the mean dopant concentration, and
$n_{sites}$ is the number of cation sites into which the dopant atom
may substitute itself.  This is easily generalized to consider
separate ``inner'' and ``outer'' volumes within the NC,
consisting of $n_{inner}$ and $n_{outer}$ sites
and having dopant concentrations $x_{inner}$ and $x_{outer}$,
respectively.  This is illustrated schematically in
\figref{inhomogeneousFig}.

The magnetization ratio, $M_{5T}/M_{sat}$ (using the notation of
Tsuji \etal), is related to the observed susceptibility, $\chi_{obs}$
and that expected of $N$ independent dopants of susceptibility
$\chi_{1}$, and can be broken down into contributions of various
dopant cluster geometries,\cite{Smart63,KMK+66} by the expression
\bea\label{magnetization_ratio}
\left ( {M_{5T}/M_{sat}} \right ) &=&
{{\frac{\chi_{obs}\mu_{0}B}{N_{d}{\chi_{1}\mu_{0}{B}}}}} \cr
& = & \sum_{i} P_{i} X_{i}
\eea
where $i$ runs over the possible cluster types, $P_{i}$ is the probability of
occurence of a given cluster of dopant atoms, and $X_{i} =
\chi_{i}/\chi_{1}$ is the normalized susceptibility of cluster-type
$i$, as calculated by Kreitman \etal\cite{KMK+66} $X_{i}$ is a
function of temperature ($T= 5{\rm K}$ in the experiment) and the
nearest-neighbor Heisenberg exchange constant, which we take to be
$-J/k_{b} \approx 10 {\rm K}$ for Mn in bulk ZnS.\cite{SFR+84} 
With (anion-mediated) nearest-neighbor interactions, one obtains from
a statistical mechanical treatment of the spin configurations as a
function of $-k_{b}T/J$ the normalized cluster suceptibilities for all
clusters up to three atoms as: $X_{single} = 1.0$ (by definition),
$X_{pair} = 0.0$, $X_{open\, triangle} = 1.0$, and $X_{closed\,
triangle} = 0.1$.\cite{KMK+66} A more detailed description is given in the Appendix.  
The values of the $X_{i}$ are
essentially constant for $5 {\rm K} \leq {-J/k_{b}} \leq 50 {\rm K}$,
so our results are independent of the exact value of $J$.

For the bulk case, the probabilities of occurence for the various
clusters have been derived by Behringer.\cite{Behr58} For nanocrystalline
systems, this is unsatisfactory, due to the finite size and shape
effects on the probability of forming various clusters; recently,
Suyver \etal\, have performed numerical simulations and derived
analytic results for the probability of dopant pairs in spherical
NCs assuming homogeneous doping.\cite{SMKM03} Less than 1\% of Mn atoms are part of
three-dopant clusters for the concentrations in our calculations, so
the pair-only truncation would make a minor contribution to the
magnetization ratio.  However, we have included them in our
calculations, as the determination is not particularly difficult.
More important is the possibility of an inhomogeneous
distribution of dopant atoms within the NC, illustrated
schematically in \figref{inhomogeneousFig}.  We performed
numerical simulations, in which \eqnref{prob_law} was used to
determine a number of dopant atoms to be randomly placed within a
given set of cation shells.  $P_{i}$ was evaluated by counting all the
types of (anion-mediated) nearest-neighbor clusters in a statistical
sampling of $10^{6}$ dopant configurations.  Then
\eqnref{magnetization_ratio} was used to determine the magnetization
ratio for the ensemble of doped NCs.

\section{Results}

Using the model of a 729 atom, 3.0 nm ZnS NC, we varied
the local dopant concentration, keeping the
concentration over the total NC equivalent to the
magnetically determined concentration, ${\bar{x}}_{curie}$, from
experiment.  The shells are explicitly described in \tblref{shells},
and the results are shown in \tblref{shellCalcns}.
The higher average number of dopant atoms, $\bar{N_{d}}$, 
in our calculations reflects the fact that
our model NC is slightly larger than that assumed by Tsuji
\etal; otherwise we would expect ${\bar{x}}_{curie}$ to
underestimate the number of dopants by a few percent.\cite{SFR+84}

We initially performed calculations in which the dopant
atoms were uniformly distributed over all 420 cation sites; 
this results in a higher magnetization ratio
for the low-concentration samples than for the high-concentration
samples, contradicting the experimental results.  However, we
understand the acid washing procedure as removing any dopant atoms on
the outermost shell of the NC.  The $x_{NC}$ reported for the
``HCl-washed'' experimental samples corresponds to the number of dopant atoms
in the interior divided by the total number of cation sites,
$n_{total}$, in the NC (including the surface
cation sites).  But in the ``HCl-washed'' case, the remaining dopants
are not in the outermost shell, but are constrained to be in the
$n_{inner}$; we can determine $x_{inner}$ using the relation
$x_{NC} n_{total} = x_{inner}
n_{inner}$.  Similar stoichiometric reasoning allows us to
calculate $x_{outer}$.  The calculations in \tblref{shellCalcns} show
the results of constraining the inner sites to consist of the
one, two, three, and four cation shells, both with and without doping
into the exterior (fifth) cation shell.  For cases with an outer
shell and higher $x_{NC}$ (modelling the ``as prepared'' experimental
samples), $x_{outer} > x_{inner}$ reduces the probability of pairing
between surface and interior dopant atoms, increasing $M_{5T}/M_{sat}$
as compared to the uniformly distributed case.  However, the
magnetization ratio for these high $x_{NC}$ samples is relatively
insensitive to any restriction of the sites available for the interior
dopant atoms, since the number of dopant atoms in this exterior shell
is much higher than the number of dopant atoms in the interior, and to
a first approximation is similar to the result obtaining by having all
the dopant atoms on the exterior (since $x_{outer}n_{outer}
\gg x_{inner}n_{inner}$).

Constraining the interior dopant atoms to reside within some locally
concentrated sphere (for the purpose of our model, concentric with the
NC), is qualitatively sufficient to achieve the proper effect
of reducing the magnetization ratio for the washed samples with
respect to the unwashed samples (as seen by comparing the respective
rows of the ``With Exterior Shell'' and ``Without Exterior Shell''
calculations in \tblref{shellCalcns}).  In particular, constraining the
interior dopant atoms to reside on the innermost cation shell nearly
reproduces the experimental magnetization ratio results for both the
``as prepared'' (93.4\% theory, 91\% experiment) and the
``HCl-washed'' (80.1\% theory, 77\% experiment), despite the
simplicity of the model.

\section{Conclusion}

The inhomogeneous dopant distribution model reproduces the
magnetization behavior observed by Tsuji \etal without
assuming different magnetic interactions
for surface and interior dopant atoms. As noted in \sectref{model}, at
the experimental temperature of $5 {\rm K}$, Heisenberg exchange
constants in the range of $5 {\rm K} \leq {-J/k_{b}} \leq 50 {\rm K}$
are indistinguishable.\cite{KMK+66} Further study on the effects of
dopant geometry and NC size on $J$ may rely on
electronic structure calculations, such as the recently developed
ZILSH method of O'Brien and Davidson.\cite{OD03}

Our results suggests that dopant atoms in the interior of the NC are
localized.  It has been noted previously that $\rm{Mn_{x}Cd_{1-x}Se}$
NCs appear to effectively zone-refine out the dopant atoms, since the
energetic barrier to rearranging the lattice is not as insurmountable
as in a bulk crystal.\cite{MKB+00} This suggest that there may exist a
region near the center of the NC in which the energetic barrier is
high enough to prevent migration of dopant atoms to the surface,
surrounded by a region in which dopants migrate to the surface during
synthesis and processing, as shown schematically in
\figref{inhomogeneousFig}.  This may be further investigated by
molecular dynamics simulations.

\section{Acknowledgements}
J.S. thanks the National
Defense Science and Engineering Grant (NDSEG) program and U.S. Army
Research Office Contract/Grant No. FDDAAD19-01-1-0612 for financial
support.  K.B.W. thanks the Miller Institute for Basic Research in
Science for financial support.  This work was also supported by the
Defense Advanced Research Projects Agency (DARPA) and the Office of
Naval Research under Grant No. FDN00014-01-1-0826.

\section{Appendix}
Below is a brief summary of the cluster susceptibility model, following
Kreitman \etal\cite{KMK+66} and Smart\cite{Smart63}.   The Heisenberg
Hamiltonian for nearest-neighbor interactions is
\be\label{hamiltonian}
\hat{H} = -2 J\sum_{nn(i,j)}{\bf S}_{i} \cdot {\bf S}_{j} + g \mu_{B}
H_{0}\sum_{i}S_{i}^{z},
\ee
where $J$ is the exchange constant, $nn(i,j)$ indicates that only
nearest-neighbor $i$ and $j$ pairs are included, ${\bf S}_{i}$ is the
spin-vector for site $i$, $g$ is the $g$-factor, $\mu_{B}$ is the Bohr
magneton, $H_{0}$ is the applied field, and $S_{i}^{z}$ is the $z$
component of the spin on site $i$.  There are three types of
nearest-neighbor clusters containing three atoms or less: pairs ($D$),
closed triangles ($E$) and open triangles ($N$).  The eigenenergies of
\eqnref{hamiltonian} for these configuratations are respectively,
\bea
E_{D} &=& -J [ j(j+1)-2S(S+1) ]+g\mu_{B}H_{0}M \cr
&  & 0 \leq j \leq 2S, \vert
M\vert\leq j \cr
E_{E} & = & -J[j(j+1)-2S(S+1)]+g\mu_{B}H_{0}M \cr
& & 0\leq k \leq 2S, \vert
k - S\vert \leq j \leq k+S,\vert M \vert \leq j \cr
E_{N} & = & - J[j(j+1)-k(k+1)-S(S+1)]+g\mu_{B}H_{0}M \cr 
&  &  0 \leq k \leq
2S, \vert
k - S\vert \leq j \leq k+S,\vert M \vert \leq j, 
\eea
using the standard notation of spin angular momentum.  For $n\in
\lbrace D,E,N \rbrace$, the free energy, $F$, is given by,
\be
- \beta F = \ln \sum_{n}e^{-\beta E_{n}}
\ee
where $\beta = 1/k_{B}T$, $k_{B}$ is the Boltzmann constant and $T$ is
the temperature.  The susceptibility, $\chi$, is given by
\bea
\chi & = & - \left . \left ( {\frac{{\partial}^{2}F}{\partial
H_{0}^{2}}}\right ) \right\vert_{H_{0}=0} \\
& = & - \left . {\frac{\partial}{\partial H_{0}}} {\frac{g \mu_{B}
\sum_{n}{\left(\sum_{i}S_{i}^{z}\right )}^{2} e^{-\beta
E_{n}}}{\sum_{n}e^{-\beta E_{n}}}} \right \vert_{H_{0}=0} \\
 & = & \left [ {\frac{\beta {(g \mu_{B})^{2}}
\sum_{n}{\left(\sum_{i}S_{i}^{z}\right )}^{2} e^{-\beta
E_{n}}}{\sum_{n}e^{-\beta E_{n}}} - {\left ( {\frac{g \mu_{B}
\sum_{n}{\left(\sum_{i}S_{i}^{z}\right )}^{2} e^{-\beta
E_{n}}}{\sum_{n}e^{-\beta E_{n}}}}\right )}^{2} }\right ]_{H_{0}=0} \\
& = & \beta (g \mu_{B})^{2} { \left [ \langle {(\sum_{i}S_{i}^{z} )}^{2} \rangle - 
{\langle\sum_{i}S_{i}^{z} \rangle} ^{2} \right]}_{H_{0}=0}. \\
\eea
The susceptibility of a set of noninteracting ($J=0$) spins,
$\chi_{S}$, is simply,
\be
\chi_{S} = {\frac{1}{3}}\beta {(g\mu_{B})}^{2}S(S+1).
\ee
We can then use this to definte a normalized susceptibility $X =
\chi/{\chi}_{S}$, and likewise normalized susceptibilities of each cluster
type $i$, as $X_{i}= {\chi_{i}}/{\chi}_{S}$, used in
\eqnref{magnetization_ratio}.  

%\bibliography{Mn_DMS}
%\bibliographystyle{apsrev}

\vfill\eject
\renewcommand{\baselinestretch}{2}

\iffalse
\begin{table}
\begin{tabular}{c c c}\hline
Type & Symbol & $X_{i}$ \\
\hline
Single & S & 1.0 \\
Pair  & D & 0.0 \\
Open triangle & N & 1.0 \\
Closed triangle & E & 0.1 \\
\hline
\end{tabular}
\caption{Nearest-neighbor cluster types and normalized
susceptibilities, following Kreitman \etal\cite{KMK+66}  We use the estimate of
$-J/k_{b} \approx 10 {\rm K}$ for Mn in bulk ZnS made by Shapira
\etal,\cite{SFR+84} and a temperature of $5 {\rm K}$ to match the
experimental details of Tsuji \etal\cite{TKK03} However, these
values stay essentially constant over the range of $ 0.1 \leq
k_{b}T/{\vert J \vert} \leq 1.0$.}
\label{clusterRef}
\end{table}
\fi

\vfill\eject
\begin{table}
\begin{tabular}{c c c c c}\hline
Shell & Type & Atoms in Shell & Total Cations & Total Atoms \\
\hline 
0 & Anion & 1 & & 1 \\
1 & Cation & 4 & 4 & 5 \\
2 & Anion & 12 & & 17 \\
3 & Cation & 24 & 28 &41 \\
4 & Anion & 42 & & 83 \\
5 & Cation & 64 & 92 &147 \\
6 & Anion & 92 & &239 \\
7 & Cation & 124 & 216 &363 \\
8 & Anion & 162 & &525 \\
9 & Cation & 204 & 420 &729 \\
\hline
\end{tabular}
\caption{Description of the zinc-blende NC model of Lippens and Lanoo \cite{LL89} for a 
3.0-nm diameter ZnS crystal.}
\label{shells}
\end{table}

\vfill\eject
\begin{table}
\begin{tabular}{c c c c c c c}\hline
$n_{inner}$ & $x_{inner}\% $ & $n_{outer}$ & $x_{outer}\% $ & $x_{NC}\%$ & 
$\bar{N_{d}}$ & $M_{5T}/M_{sat} \%$ \\
\hline
\multicolumn{7}{l}{Uniformly Distributed}\\
 420  &  1.20  &     0  &  0.00  &  1.20  &   5.03  &     90.3 \\
 420  &  0.10  &     0  &  0.00  &  0.10  &   0.42  &     99.1 \\
\multicolumn{7}{l}{With Exterior Shell} \\ 
 216  &  0.19  &   204  &  2.27  &  1.20  &   5.02  &     93.7 \\
  92  &  0.46  &   204  &  2.27  &  1.19  &   5.01  &     94.5 \\
  28  &  1.50  &   204  &  2.27  &  1.19  &   5.00  &     94.0 \\
   4  &  10.48  &   204  &  2.27  &  1.17  &   4.91  &     93.4 \\
\multicolumn{7}{l}{Without Exterior Shell} \\
 216  &  0.19  &   204  &  0.00  &  0.10  &   0.42  &     98.2 \\
  92  &  0.46  &   204  &  0.00  &  0.10  &   0.41  &     96.2 \\
  28  &  1.50  &   204  &  0.00  &  0.10  &   0.40  &     90.6 \\
   4  &  10.48  &   204  &  0.00  &  0.07  &   0.31  &     80.1 \\
\multicolumn{7}{l}{Exterior Shell Only} \\
   0  &  0.00  &   204  &  2.27  &  1.09  &   4.60  &     94.3 \\
\multicolumn{7}{l}{All Dopants on Exterior}\\
   0  &  0.00  &   204  &  2.47  &  1.20  &   5.02  &     93.8 \\
   0  &  0.00  &   204  &  0.21  &  0.10  &   0.42  &     99.5 \\
\hline
\multicolumn{7}{l}{Experimental Results\cite{TKK03}} \\
\multicolumn{4}{c}{As prepared} & 1.2 & 4.3 & 91 \\
\multicolumn{4}{c}{HCl-washed} & 0.10 & 0.3-0.5 & 77 \\
\hline
\end{tabular}

\caption{Calculated results, for $10^{6}$ NC sample.  Definitions of
the columns are given in Sections 2 and 3.  We found the values of
$M_{5T}/M_{sat}$ and $\bar{N_{d}}$ to vary only slighlty as compared
to calculations using $10^{4}$ NCs.}

\label{shellCalcns}
\end{table}

\end{document}